# Robust Convergence of Power Flow using Tx Stepping Method with Equivalent Circuit Formulation

Amritanshu Pandey[1], Marko Jereminov[1], Martin R. Wagner[1], Gabriela Hug[2,1], Larry Pileggi[1]

[1]Dept. of Electrical and Computer Engineering
Carnegie Mellon University
Pittsburgh, PA

[2]Power Systems Laboratory
ETH Zurich
Switzerland

*Abstract –* **Robust solving of critical large power flow cases (with 50k or greater buses) forms the backbone of planning and operation of any large connected power grid. At present, reliable convergence with applications of existing power flow tools to large power systems is contingent upon a good initial guess for the system state. To enable robust convergence for large scale systems starting with an arbitrary initial guess, we extend our equivalent circuit formulation for power flow analysis to include a novel continuation method based on transmission line ('Tx') stepping. While various continuation methods have been proposed for use with the traditional 'PQV' power flow formulation, these methods have either failed to completely solve the problem or have resulted in convergence to a low voltage solution. The proposed "Tx Stepping" method in this paper demonstrates robust convergence to the high voltage solution from an arbitrary initial guess. Example systems, including 75k+ bus test cases representing different loading and operating conditions for Eastern Interconnection of the U.S. power grid, are solved from arbitrary initial guesses.**

*Index Terms—* **continuation methods/homotopy, equivalent circuit formulation, power flow, robust convergence, Tx Stepping**

## I. Introduction

An accurate solution to the power flow problem is essential for secure operation and planning of the power grid. The industry standard for solving the power flow problem is based on the 'PQV' formulation [1], wherein nonlinear power mismatch equations are solved for bus voltage magnitude and angle state variables that define the steady-state operating point of the system. However, this formulation is characterized by highly nonlinear power balance equations that are known to suffer from lack of robustness [2], particularly with the increase in the scale of the system. Of the many known challenges contributing toward lack of robustness, the two that are the most detrimental are: i) convergence to non-physical solution [3] and ii) divergence [2].

The factors that are the most fundamental toward making the power flow problem challenging are i) the use of non-physical representations for modeling the power grid components and ii) the use of power mismatch equations with real and reactive power as system state variables to formulate the problem. The non-physical models, such as the PV model for the generator, can result in convergence to a non-physical solution or divergence. Similarly, non-linearities in the 'PQV' formulation almost always cause divergence for large (>50k) and ill-conditioned test cases when solved using an arbitrary (e.g. flat start) set of initial conditions. This lack of a physics based formulation along with methods that can constrain the non-physics based models in their physical space is what renders the existing power flow problem and solution approaches to be "non-robust".

In order to develop a robust power flow solver, it is imperative that the solver can efficiently and effectively navigate through these challenges while converging to a solution that is both meaningful and correct. We define a solution as being the "correct physical solution" if the system voltages are within the acceptable range and the angle differences between connected adjacent nodes are less than 90°. To achieve this we propose a two pronged approach: i) the use of equivalent circuit formulation with true state variables of currents and voltages [4]-[6] to model the grid and ii) the use of circuit simulation methods to ensure robust convergence to correct physical solutions. In [7], we showed that with the use of circuit simulation methods applied to an equivalent circuit of the considered power grid, we can robustly converge for test cases up to 15k buses to correct physical solution from arbitrary initial guesses. However, for large power flow cases (> 50k+ buses) with complex models (such as remote voltage control, FACTS devices etc.), more extensive management of convergence is sometimes required.

In this paper, we propose a homotopy continuation method that we refer to as "Tx Stepping" to achieve robust convergence for large scale power flow cases from a set of arbitrary initial guesses. Homotopy methods have been previously studied for the power flow problem [2], [8] but have mostly been unsuccessful due to convergence to low voltage solutions or their inability to scale to large cases [9]. Tx stepping is based on the physics of the grid and takes inspiration from the "gmin stepping" method in the circuit simulation domain. In the gmin stepping method, all the nodes in the circuit model are initially shunted/shorted to ground such that a trivial solution to the problem exists. Analogously, in the proposed Tx stepping method for power systems, we "virtually short" the transmission lines and transformers in order to obtain a trivial initial solution. The system is then gradually relaxed until the original problem is solved. Furthermore, the method also modifies the problem such that the non-linearities due to remote voltage control, phase shifters etc., are all relaxed initially such that a trivial solution exists for the problem independent of its size.

The remainder of the paper is structured as follows, In Section II we provide a brief background of equivalent circuit formulation and previously proposed circuit simulation methods for the power flow problem. Sections III and IV discuss the homotopy methods and the novel Tx stepping method in detail. The implementation of the Tx stepping method in equivalent circuit formulation and the subsequent advantages of the approach are discussed in Section V. In the result section, we introduce our solver SUGAR (Simulation with Unified Grid Analyses and Renewables) and apply the Tx stepping method to multiple large test systems with 75k+ nodes and demonstrate robust convergence to correct physical solutions from a set of arbitrary initial guesses with no prior knowledge of the system state.

## II. BACKGROUND

### A. Equivalent Circuit Formulation

The equivalent circuit approach for generalized modeling of the power system in steady-state (i.e. power flow and three-phase power flow) was recently introduced in [4]-[6]. This circuit-based formulation represents both the transmission and distribution power grid in terms of equivalent circuit elements. It was shown that each of the power system components (including constant power models, i.e. PQ and PV buses) can be directly mapped to an equivalent circuit model based on the underlying relationship between current and voltage state variables without loss of generality. Importantly, this formulation can represent any physics based load model or measurement based semi-empirical models as a sub-circuit that can then be combined hierarchically with other circuit abstractions to build larger aggregated models. The equivalent circuit representations of the most prominent models for the power flow problem are summarized in the following sections.

#### 1) PV Bus

The equivalent circuit formulation provides a choice to model the constant voltage (PV) node as either a complex voltage source (as functions of complex current) [5] or a complex current source (as functions of complex voltage) [4]. It has been shown that representing the PV bus as a complex current source offers superior convergence when applying Newton-Raphson (NR) iterations to the resulting equation system. To enable the application of NR, the complex current source is split into real and imaginary current sources ($I_{RG}$ and $I_{IG}$, respectively). This is necessary due to the non-analyticity of complex conjugate functions [4]. The resulting equations are:

$$I_{RG} = \frac{P_G V_{RG} + Q_G V_{IG}}{V_{RG}^2 + V_{IG}^2} \quad (1)$$

$$I_{IG} = \frac{P_G V_{IG} - Q_G V_{RG}}{V_{RG}^2 + V_{IG}^2} \quad (2)$$

An additional constraint that allows the generator to control the voltage magnitude either at its own node or any other remote node in the system is represented by a control circuit, as shown in the following subsection. The reactive power $Q_G$ of the generator acts as the additional unknown variable for the additional constraint that is introduced due to voltage control.

The first order terms of the Taylor expansions for (1) and (2) are used to linearize the functions and derive an equivalent circuit model, as shown in Fig. 1. For example, linearization of the real generator current is:

$$I_{RG}^{k+1} = \frac{\partial I_{RG}}{\partial Q_G}|_{Q_G^k, V_{RG}^k, V_{IG}^k}(Q_G^{k+1}) + \frac{\partial I_{RG}}{\partial V_{RG}}|_{Q_G^k, V_{RG}^k, V_{IG}^k}(V_{RG}^{k+1})$$
$$+ \frac{\partial I_{RG}}{\partial V_{IG}}|_{Q_G^k, V_{RG}^k, V_{IG}^k}(V_{IG}^{k+1}) + I_{RG}^k - \frac{\partial I_{RG}}{\partial Q_G}|_{Q_G^k, V_{RG}^k, V_{IG}^k}(Q_G^k) \quad (3)$$
$$- \frac{\partial I_{RG}}{\partial V_{RG}}|_{Q_G^k, V_{RG}^k, V_{IG}^k}(V_{RG}^k) - \frac{\partial I_{RG}}{\partial V_{IG}}|_{Q_G^k, V_{RG}^k, V_{IG}^k}(V_{IG}^k)$$

The first term in (3) represents a current source that is a function of the reactive power; the second term represents a conductance, since the real current is proportional to the real voltage; the third term represents a voltage-controlled current source, since the real current is proportional to the imaginary voltage. The remaining terms are all dependent on known values from the previous iteration, so they can be lumped together and represented as an independent current source.

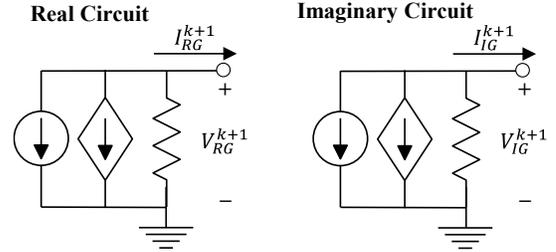

*Figure 1: Equivalent Circuit Model for PV generator model.*

#### 2) Voltage Regulation of the Bus

Numerous power grid elements such as generators, FACTS devices, transformers, shunts etc., are capable of controlling a voltage magnitude at a given node in the system. Moreover, they can control the voltage magnitude at either their own node $\mathcal{O}$ or a remote node $\mathcal{W}$ in the system. In equivalent circuit formulation, we represent the control of the voltage magnitude by a control circuit (Fig. 2) governed by

$$F_\mathcal{W} \equiv V_{set}^2 - V_{R\mathcal{W}}^2 - V_{I\mathcal{W}}^2 = 0 \quad (4)$$

The circuit in Fig. 2 is derived from the linearized version of (4). It is stamped for each node $\mathcal{W}$ in the system whose voltage is being controlled such that there exists at least one single path between the node $\mathcal{W}$ and the equipment's node $\mathcal{O}$ that is controlling it. The additional unknown variable for this additional constraint is dependent on the power system device that is controlling the voltage magnitude. For example, the additional unknown variable for a generator is its reactive power $Q$, whereas in the case of transformers, it can be the transformer tap $tr$, and for FACTS devices it can be the firing angle $\varphi$. The previous section showed how the additional unknown variable for PV buses is integrated in the respective equivalent circuits for generators.

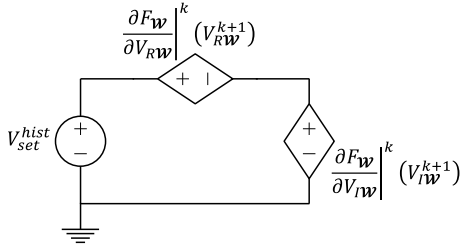

Figure 2: Voltage magnitude constraint control equivalent circuit.

*3) PQ Bus*

Similar to the PV bus, the constant power node (PQ bus) is also represented as an equivalent circuit via either a complex voltage source or a complex current source. It has been empirically determined that superior convergence is observed when the load bus is modeled as complex current source. The two fundamental equations for the real and imaginary currents for the PQ buses are given by:

$$I_{RL} = \frac{P_L V_{RL} + Q_L V_{IL}}{V_{RL}^2 + V_{IL}^2} \quad (5)$$

$$I_{IL} = \frac{P_L V_{IL} - Q_L V_{RL}}{V_{RL}^2 + V_{IL}^2} \quad (6)$$

Linearizing the load model in (5) and (6) via Taylor expansion results in three elements in parallel for both real and imaginary circuits: a conductance, a voltage-controlled current source, and an independent current source.

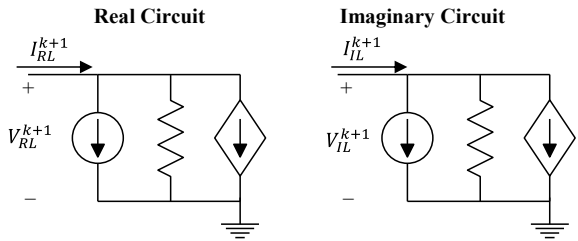

Figure 3: Equivalent split-circuit PQ load model.

*4) Physics Based Models*

It has been previously shown in [10] that any physics based device model can also be directly incorporated into the equivalent circuit formulation. For instance, consider the three-phase induction motor (IM) example described in [10]. The steady state and transient behavior of an IM can be expressed by a set of five ordinary differential equations. These mathematical expressions are mapped into an equivalent circuit (as shown in Fig. 4) using standard circuit simulation techniques [11]. Due to the use of DQ transformation [10], this physics based equivalent circuit model of an IM can be directly used for steady state power flow formulations by shorting the inductors and opening (open-circuiting) the capacitors.

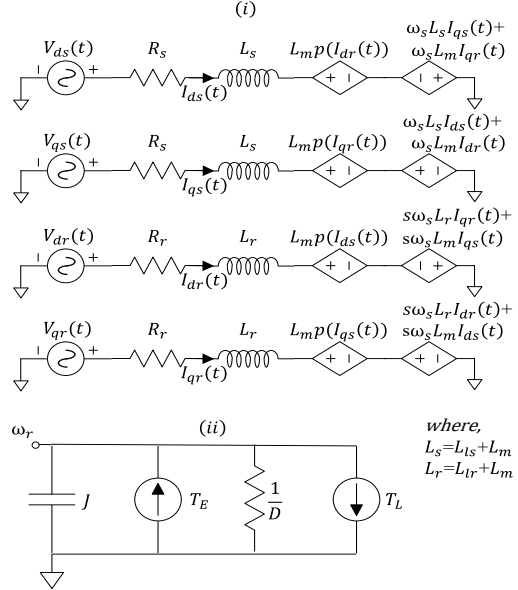

Figure 4: Equivalent circuit for the three-phase induction motor model in natural state variables of I-V.

*5) BIG Model*

The BIG aggregated load model introduced in [12]-[13] (Fig. 5) was shown to more accurately capture the actual load behavior when compared with more traditional load models such as ZIP, PQ, etc., and can be easily derived from real-time measurement data. Importantly, the BIG load model is linear in equivalent current/voltage split circuit formulation, hence it results in linear equality constraints for the load bus in the power-flow analysis.

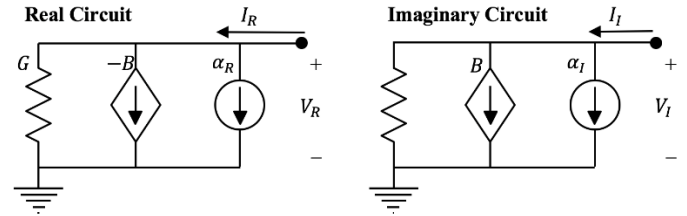

Figure 5: Equivalent circuit of a BIG load model.

*B. Circuit Simulation Techniques*

Decades of research in circuit simulation have demonstrated that circuit simulation methods can be applied for determining the DC state of a highly non-linear circuits using NR. These techniques have been shown to make NR robust and practical for large-scale circuit problems [14] consisting of billions of nodes. Most notable is the ability to guarantee convergence to the correct physical solution (i.e. global convergence) and the capability of finding multiple operating points [14]. We have previously proposed analogous techniques for ensuring convergence to the correct physical solution for the power flow and three-phase power flow problems [7]. In this section, we provide a short overview over these techniques.

*1) Variable Limiting*

The solution space of the system node voltages in a power flow problem is well defined. While solving the power flow problem, a large NR step may step out of this solution space and result in either divergence or convergence to a non-physical or incorrect solution. It is, therefore, important to limit the NR step before an invalid step out of the solution space is made. In [7] we proposed variable limiting to achieve the postulated goal. In this technique, the state variables that are most sensitive to initial guesses are damped when the NR algorithm takes a large step out of the pre-defined solution space. Note however, that *not all* of the system variables are damped for the variable limiting technique, as is done for traditional damped NR. The circuit simulation research has shown that damping most sensitive variables provides superior convergence compared to damped NR in general.

The plot in Fig. 6 shows results of variable limiting for a 2383 bus test system for which the equivalent circuit was formulated using I-V variables. Simulations were run for six different initial guesses using unspecified Q (reactive power) supplied by the generators. The maximum bus voltage from the solution of the power flow problem for each initial guess is shown for two scenarios: without and with variable limiting enabled. The plots show that when variable limiting is not enabled, in most cases the voltage solution diverges to very high magnitudes (up to $10^4$). However, when the variable limiting option is enabled, divergence is not observed and the bounded bus voltages result in fast convergence.

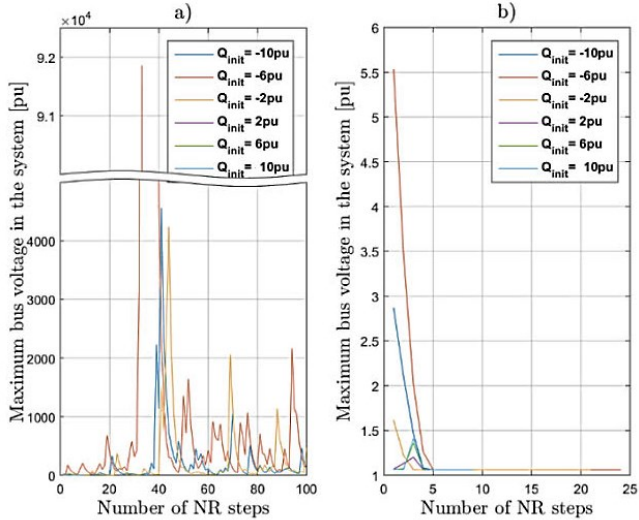

*Figure 6: Voltage profile for maximum bus voltage in 2383 Bus System: a) w/o Variable Limiting b) with Variable Limiting*

In order to apply variable limiting in our prototype simulator, the mathematical expressions for the PV nodes in the system are modified as follows:

$$I_{CG}^{k+1} = \varsigma \frac{\partial I_{CG}}{\partial V_{RG}}(V_{RG}^{k+1} - V_{RG}^k) + I_{CG}^k \\ + \varsigma \frac{\partial I_{CG}}{\partial V_{IG}}(V_{IG}^{k+1} - V_{IG}^k) + \frac{\partial I_{CG}}{\partial Q_G}(Q_G^{k+1} - Q_G^k) \quad (7)$$

where, $0 \leq \varsigma \leq 1$ and $C \in \{R, I\}$ represents the placeholder for real and imaginary parts. The magnitude of $\varsigma$ is dynamically varied through heuristics such that convergence to the correct physical solution is achieved in the most efficient manner. The heuristics depend on the largest delta voltage ($\Delta V_R, \Delta V_I$) step during subsequent NR iterations. If during subsequent NR iterations, a large step ($\Delta V_R, \Delta V_I$) is encountered, then the factor $\varsigma$ is decreased. The factor $\varsigma$ is scaled back up if consecutive NR steps result in monotonically decreasing absolute values for the largest error.

*2) Voltage Limiting*

An equally simple yet effective technique is to limit the absolute value of the delta step that the real and imaginary voltage vectors are allowed to make during each NR iteration. This is analogous to the voltage limiting technique used for diodes in the circuit simulation wherein the maximum allowable voltage step during NR is limited to twice the thermal voltage of the diode. Furthermore, for power flow analysis based on the equivalent circuit formulation, a hard limit is enforced on the real and imaginary voltages in the system. The mathematical implementation of voltage limiting in our formulation is as follows:

$$V_C^{k+1} = \min_{V_C^{min}} \max_{V_C^{max}} (V_C^k + \delta_S \min(|\Delta V_C^k|, \Delta V_C^{max})) \quad (8)$$

where $\delta_S = sign(\Delta V_C^k)$ and $C \in \{R, I\}$ represents the placeholder for real and imaginary parts

### III. HOMOTOPY METHOD

In our previous publications [7], circuit simulation methods with equivalent circuit formulation were shown to achieve robust convergence for the power flow problems from an arbitrary set of initial conditions for test cases up to 15k buses. However, for systems that are greater than 50k buses, these methods alone are at times unable to achieve convergence to correct physical solution from an arbitrary set of initial conditions. For these cases, the solver either converged to a low voltage solution or diverged altogether.

To tackle these challenges we propose the use of homotopy methods to ensure convergence for the system to the correct physical solution independent of its complexity or scale. Homotopy methods that have been proposed in the past [2], [8] have suffered from convergence to low voltage solutions [2] and divergence. Furthermore, none of the previously proposed homotopy methods are known to scale up to test systems [9] that are of the scale of European or the US grids, which is essential for secure operation and operation of these systems. For example, the Eastern Interconnection system of the US grid is composed of more than 80,000 buses.

With homotopy methods, the original problem is replaced with a set of sub-problems that are sequentially solved. The set of sub-problems exhibit certain properties: i) the first sub-problem has a trivial solution and ii) each incrementally subsequent problem has a solution very close to the solution of the prior sub-problem. Mathematically this can be described via the following expression:

$$\mathcal{H}(x, \lambda) = (1 - \lambda)\mathcal{F}(x) + \lambda \mathcal{G}(x) \quad (9)$$

where $\lambda \in [0, 1]$.

The method begins by replacing the original problem $\mathcal{F}(x) = 0$ with $\mathcal{H}(x, \lambda) = 0$. The equation set $\mathcal{G}(x)$ is a representation of the system that has a trivial solution. The homotopy factor $\lambda$ has the value of 1 for the first sub-problem and therefore the initial solution is equal to trivial solution of $\mathcal{G}(x)$. For the final sub-problem that corresponds to the original problem, the homotopy factor $\lambda$ has the value of zero. In order to generate sequential sub-problems, the homotopy factor is dynamically decreased in small steps until it has reached the value of zero.

## IV. Tx (Transmission Line) Stepping

We propose a new homotopy approach "Tx Stepping" that is specifically defined for the non-linearities observed in the power flow and three-phase power flow problems.

### A. General Approach

The series elements in the system (transmission lines, transformers etc.) are "virtually" shorted at first to solve the initial problem that has a trivial solution. Specifically, a large conductance ($G$) and a large susceptance ($B$) are added in parallel to each transmission line and transformer model in the system. Importantly, the solution to this initial problem results in high system voltages (magnitudes) as they are essentially driven by the slack bus complex voltage and the PV bus voltage magnitude due to the low voltage drops in the lines and transformers (as expected with virtually shorted systems). Similarly, the solution to bus voltage angles will lie within a $\epsilon$-small radius around the slack bus angle. Subsequently, like other continuation methods, the formulated system problem is then gradually relaxed to represent the original system by taking small increment steps of the homotopy factor ($\lambda$) until convergence to the solution of the original problem is achieved. Mathematically, this is expressed by:

$$\forall i \in \{Tx, Xfmrs\} : \hat{G}_i = G_i + \lambda\gamma G_i \quad (10)$$

$$\forall i \in \{Tx, Xfmrs\} : \hat{B}_i = B_i + \lambda\gamma B_i \quad (11)$$

$$\forall i \in Shunts : \hat{G}_i^{sh} + j\hat{B}_i^{sh} = (1 - \lambda\gamma)(G_i^{sh} + jB_i^{sh}) \quad (12)$$

where, $Xfmrs$ is the set of all transformers and $Tx$ is the set of all the transmission lines in the system. $G_i$, $B_i$, $G_i^{sh}$ and $B_i^{sh}$ are the original system impedances and the $\hat{G}$, $\hat{B}$, $\hat{G}^{sh}$ and $\hat{B}_i^{sh}$ are the system impedances used while iterating from trivial problem to the original problem. The parameter $\gamma$ is used as a scaling factor for the conductances ($G$) and susceptances ($B$). If the homotopy factor ($\lambda$) takes the value one, the system has a trivial solution and if its takes the value zero, the original system is represented.

Along with ensuring convergence for a problem, Tx stepping avoids the undesirable low voltage solutions for the power flow problem since the initial problem results in a solution with high system voltages, and subsequent step of the homotopy approach continues and deviates ever so slightly from this initial solution, thereby guaranteeing convergence to the high voltage solution for the original problem.

### B. Handling of Transformer Phase Shifters and Taps

In order to "virtually short" a power system, we must also account for transformer taps $tr$ and phase shifters $\theta$. In a "virtually" shorted condition, all the nodes in the system have complex voltages that is in close proximity to the slack bus or PV bus complex voltages, which can be intuitively defined by a small epsilon norm ball around these voltages. Therefore, in order to achieve the following form we must modify the transformer taps and phase shifter angles such that at $\lambda = 1$, their turns ratios and phase shift angles correspond to a magnitude of 1 pu and 0°, respectively. Subsequently, the homotopy factor $\lambda$ is varied such that the original problem is solved with original transformer tap and phase shifter settings. This can be mathematically expressed as follows:

$$\forall i \in Xfmrs : \hat{tr}_i = tr_i + \lambda(1 - tr_i) \quad (13)$$

$$\forall i \in Xfmrs : \hat{\theta}_i = \theta_i - \lambda\theta_i \quad (14)$$

### C. Handling of Voltage Control for Remote Buses

To achieve a trivial solution during first step of Tx stepping it is essential that we also handle remote voltage control appropriately. Remote voltage control refers to a phenomenon wherein a device on node $\mathcal{O}$ in the system can control the voltage of another node $\mathcal{W}$ in the system. This behavior is highly non-linear and if not handled correctly can result in divergence or converge to low voltage solution. Existing commercial tools for power flow suffer from this problem and lack the robustness to handle remote voltage control effectively. Therefore, in the Tx stepping method, we introduce a technique wherein we "virtually short" the path between the controlling node ($\mathcal{O}$) and the controlled node ($\mathcal{W}$) at $\lambda = 1$, such that the device at the controlling node can easily supply the currents to the controlled node $\mathcal{W}$ and control its voltage. Subsequently during homotopy, we gradually relax the system such that additional line connecting the controlling node ($\mathcal{O}$) and controlled node ($\mathcal{W}$) is open at $\lambda = 0$.

## V. Implementation of Tx Stepping in Equivalent Circuit Formulation

Unlike traditional implementations of homotopy methods, in our equivalent circuit formulation we do not directly modify the non-linear set of mathematical equations but instead embed a homotopy factor in each of the equivalent circuit models for the power grid components. In doing so we allow for incorporation of any power system equipment into the Tx stepping approach within the equivalent circuit formulation framework without loss of generality. Furthermore, we ensure, that the physics of the system is preserved while modifying it for the homotopy method. Fig. 7 and 8 represent the equivalent circuits for transmission lines and transformers respectively, with homotopy factor λ embedded in them. The value of λ lies in the closed set [0, 1] and can take any value in the set.

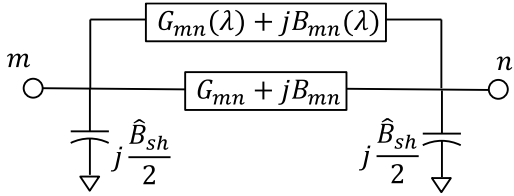
Figure 7: Homotopy factor embedded in transmission line equivalent circuit.

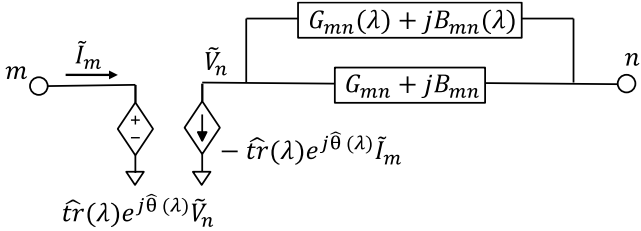
Figure 8: Homotopy factor embedded in transformer equivalent circuit.

## VI. RESULTS

Example cases were simulated in our prototype solver SUGAR (Simulation with Unified Grid Analyses and Renewables) to validate "Tx Stepping" homotopy method. The example cases include known ill-conditioned test cases and large systems that represent different operating conditions of the eastern interconnection of the US grid. We affirm that the proposed framework can guarantee convergence to correct physical solutions for all power flow cases, independent of the choice of the initial guess.

### A. Ill-Conditioned Systems

In mathematical theory, if the condition number of a given matrix is large, then the matrix and the system corresponding to that matrix are ill-conditioned. In the power flow problem, the matrix of interest is the Jacobian that is used to calculate the updated system state variables at each NR step. If the condition number of the Jacobian matrix is large at the solution point, then the system is assumed to be ill-conditioned. The 11-bus, 13-bus, and 43-bus test cases from the power system literature [15] are considered as ill-conditioned systems. However, it is systematically shown in [15] that out of these three systems, the 11-bus system is the only genuine ill-conditioned system with a maximum loading of 99.82 %. The 13-bus system is not an ill-conditioned system and can easily be solved via any power flow method, and the 43-bus test case has a maximum loading of 58 % for, which there is no feasible solution for the base loading.

TABLE 1: COMPARISON OF RESULTS FOR MODIFIED 11 BUS TEST CASE

| Initial Condition | | Ill Conditioned 11 Bus Test Case | |
|---|---|---|---|
| $V_{mag}$ (pu) | $V_{ang}$ (°) | Standard Commercial Tool[2] | SUGAR[1] |
| 1 | 0 | Low Voltage | High Voltage |
| 0.76 | 23 | Low Voltage | High Voltage |
| 0.71 | 45 | Low Voltage | High Voltage |
| High Voltage | High Voltage | High Voltage | High Voltage |

1. Tx Stepping was enabled while running simulations in SUGAR
2. Full Newton Raphson was the solver used in Standard Commercial Tool

Table 1 shows the comparison of results for the 11 bus ill-conditioned test case at 99.82 % loading for different set of initial conditions. Using standard commercial tools, for most initials conditions the system is likely to converge to a low voltage solution or diverge. Without homotopy methods, the commercial solver can only converge to the correct physical solution if the initial condition is the solution itself. However, SUGAR is able to converge to the correct physical solution from arbitrary initial conditions when Tx Stepping is applied.

Another notable ill-conditioned case is a 13659 bus system from the PEGASE test cases. At the solution point, the approximate condition number of the system Jacobian is 1.7e8. Fig. 9 shows convergence results for this test case from ten arbitrary initial conditions for a standard commercial tool and SUGAR. From the set of 10 initial conditions, the standard commercial tool converged to the correct physical solution once, diverged 8 times, and converged to the angular unstable solution one time. The ten initial conditions were chosen uniformly from the set of:

$$V_r \in [0.6, 1.1], \ V_i = \{x \in \mathbb{R}^n \mid x = 1 - V_r\}. \quad (15)$$

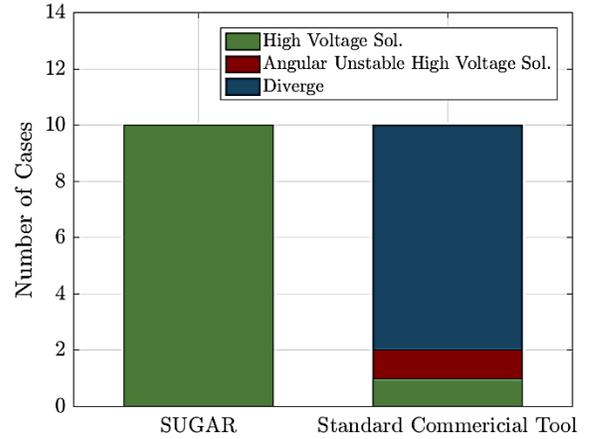
Figure 9: Results for 13659 buses PEGASE system.

### B. Large Test Cases

We next demonstrate that the Tx stepping method is scalable to large test cases and that it ensures convergence from arbitrary initial conditions. Fig. 10 shows the results for six distinct test systems that represent the eastern interconnection network of the US power grid under different loading conditions (Summer/Winter) and time periods (2017, 2018, 2021, 2026 etc.). The Tx stepping method was used to solve each of these systems from a set of different initial conditions that were uniformly chosen from the sets of:

$$V_{ang} \in [-50, 50], \ V_{mag} \in [0.6, 1]. \quad (16)$$

The vertical and horizontal axes of the figure represent the set of initial conditions ($V_{ang}, V_{mag}$) for a given case, respectively. If the case converged to a correct physical solution, it is marked via a green mark; whereas if the case diverged then it is marked via a red mark. The figure indicates that the Tx stepping method was able to achieve convergence for all of the six large eastern interconnection systems independent of the choice of initial conditions. The run time per iteration for the eastern interconnection test cases in SUGAR is comparable to other available commercial tools (less than 0.4s per iteration).

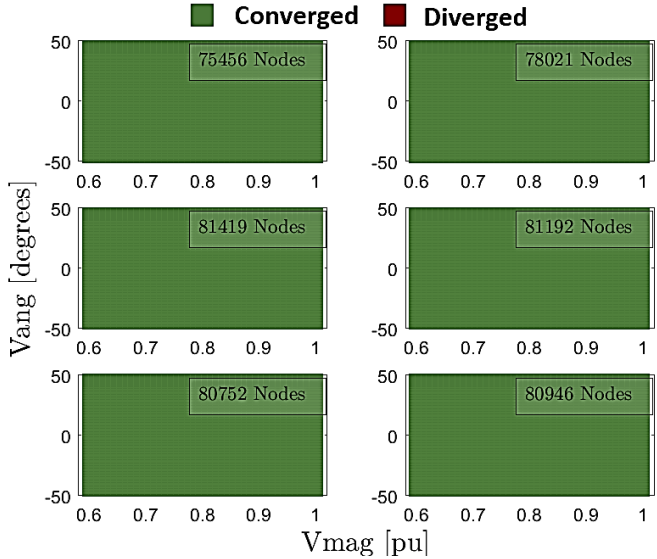

*Figure 10: Convergence sweep of large cases that represent Eastern Interconnection from range of initial conditions*

In order to further demonstrate the robustness of our approach, we consider a set of scenarios wherein we plan a realistic contingency on large test cases. The contingency in these cases is defined by loss of either two ($\mathcal{N}2$) or three ($\mathcal{N}3$) generators in the system. We then solve these cases with the contingency applied using the commercial tool and SUGAR. The initial conditions for all of the cases is chosen as the solution prior to the contingency (thereby suggesting that the system is close to its operating state post-contingency). We demonstrate the results in Table 2.

TABLE 2: CONTINGENCY ANALYSIS FOR LARGE TEST CASES

| Case | Contingency Type | No. of Buses | Standard Commercial Tool | SUGAR |
|---|---|---|---|---|
| Case 1 | $\mathcal{N}2$ | 75456 | Diverge | Converge |
| Case 2 | $\mathcal{N}2$ | 78021 | Diverge | Converge |
| Case 3 | $\mathcal{N}3$ | 80293 | Diverge | Converge |
| Case 4 | $\mathcal{N}3$ | 81238 | Diverge | Converge |

The results in Table 2 further demonstrate the robustness of our solver when using Tx Stepping. Importantly, this will enable future optimal power flow tools by being able to readily and robustly validate the security constraints on the optimal power flow dispatch and also allow for robust Monte Carlo analysis.

## VII. CONCLUSIONS

In this paper, the power flow problem is formulated using an equivalent circuit framework that when combined with novel homotopy continuation method "Tx Stepping" and other circuit simulation methods is able to achieve convergence to the correct physical solution for any test system independent of its scale or complexity. This work directly addresses known convergence issues in the existing formulations for power flow analysis and in doing so enables robust contingency analysis, security constrained optimal power flow, state estimation, and probabilistic power flow for large and ill-conditioned test cases representing complex power grid.

ACKNOWLEDGEMENTS

This work was supported in part by the Defense Advanced Research Projects Agency (DARPA) under award no. FA8750-17-1-0059 for the RADICS program.